\documentclass[final]{aipproc}

\layoutstyle{6x9}

\newcommand{\gsim}{~{}_{\textstyle\sim}^{\textstyle >}~}
\newcommand{\lsim}{~{}_{\textstyle\sim}^{\textstyle <}~}

\begin{document}

\title{Magnetic Moments of Dirac Neutrinos}


\author{Nicole F. Bell}
{address={California Institute of Technology, Pasadena, CA 91125, USA}}

\author{V. Cirigliano}
{address={California Institute of Technology, Pasadena, CA 91125, USA}}

\author{M. J. Ramsey-Musolf}
{address={California Institute of Technology, Pasadena, CA 91125, USA}}

\author{P. Vogel}
{address={California Institute of Technology, Pasadena, CA 91125, USA}}

\author{\\Mark B. Wise}
{address={California Institute of Technology, Pasadena, CA 91125, USA}}

\begin{abstract}
The existence of a neutrino magnetic moment implies contributions to the
neutrino mass via radiative corrections.  We derive model-independent
"naturalness" upper bounds on the magnetic moments of Dirac neutrinos,
generated by physics above the electroweak scale. The neutrino mass
receives a contribution from higher order operators, which are
renormalized by operators responsible for the neutrino magnetic moment.
This contribution can be calculated in a model independent way.  In the
absence of fine-tuning, we find that current neutrino mass limits imply
that $|\mu_\nu| < 10^{-14}$ Bohr magnetons.  This bound is several orders
of magnitude stronger than those obtained from solar and reactor neutrino
data and astrophysical observations.
\end{abstract}

\maketitle



Current advances in uncovering the pattern of neutrino mass and mixing,
lead naturally to questions about more exotic neutrino properties,
such as the magnetic moment, $\mu_\nu$.  In this paper, we describe how the
smallness of the neutrino mass may be used to set a strong
model-independent limit on the size $\mu_\nu$~\cite{Bell}.
Neutrino magnetic moments are reviewed in~\cite{Fukugita}, and recent
work can be found in~\cite{McLaughlin}.  In the Standard Model (SM),
extended to contain right-handed neutrinos, $\mu_\nu$ is non-zero but
unobservably small, $\mu_\nu \simeq 3
\times 10^{-19} [m_\nu/{\rm 1~eV}]$~\cite{Marciano}.  Current limits are several orders
of magnitude larger, so a magnetic moment anywhere near the present
limits would certainly be an indication of physics beyond the SM.
The best laboratory limits arise from neutrino-electron scattering.
The weak and electromagnetic contributions to $\nu-e$ scattering are
comparable if
\begin{equation}
\label{eq:muexp}
\frac{|\mu_{\nu}^{\rm exp}|}{\mu_B} 
\simeq \frac{G_F \, m_e}{\sqrt{2} \pi \alpha} \sqrt{m_e T}
\sim  10^{-10} \sqrt{\frac{T}{m_e}}\ ,
\end{equation}
where $T$ is the kinetic energy of the recoiling electron.  The
present limits derived from solar and reactor neutrino experiments are
$|\mu_\nu |\lsim 1.5 \times 10^{-10}\mu_B$~\cite{beacom} and $|\mu_\nu
|\lsim 0.9 \times 10^{-10}\mu_B$~\cite{MUNU} respectively.  A more
stringent limit can be derived from bounds on energy loss in stars,
$|\mu_\nu|\lsim 3\times 10^{-12}\mu_B$~\cite{raffelt}.

The presence of a non-zero neutrino magnetic moment will necessarily
induce a correction to the neutrino mass term.  (The problem of
reconciling a large magnetic moment with a small mass has been
recognized in the past, and possible methods of overcoming this
restriction through the use of symmetries are discussed
in~\cite{Voloshin}.)  Assuming that $\mu_\nu$ is generated by physics
beyond the SM at a scale $\Lambda$, its leading contribution to the
neutrino mass, $\delta m_\nu$, scales with $\Lambda$ as
\begin{equation}
\label{eq:leading} 
\delta m_\nu \sim \frac{\alpha}{32\pi} \frac{\Lambda^2}{m_e}
\frac{\mu_\nu}{\mu_B}\ ,  
\end{equation} 
where $\delta m_\nu$ is a contribution to $m_\nu$ arising from
radiative corrections at one-loop order. The $\Lambda^2$ dependence
arises from the quadratic divergence of the dimension four neutrino
mass operator, ${\cal{O}}^{(4)}_M
\equiv (\bar{L} \tilde{\phi}) \nu_R$.  Although the precise value of this
term cannot be calculated in a model-independent way, we can estimate
that for $\Lambda \gsim 1$ TeV and $\delta m_\nu \lsim 1~{\rm eV}$, we
require $|\mu_\nu| \lsim 10^{-14} \mu_B $.
Given the $\Lambda^2$ dependence, this bound becomes considerably more
stringent for $\Lambda$ well above the electroweak (EW) scale.
However, if $\Lambda$ is not significantly larger that the EW scale,
higher dimension operators are important, and their contribution to
$m_\nu$ can be calculated in a model independent way.

We start by constructing the most general operators that could give
rise to a magnetic moment operator, $\bar{\nu}_L \sigma^{\mu\nu}
F_{\mu\nu} \nu_R$.  Demanding invariance under the SM gauge group
$\rm{SU}(2)_L \times \rm{U}(1)_Y$, we have the following 6D operators
%
%
\begin{equation}
\label{eq:ops} 
{\cal O}^{(6)}_B  =  \frac{g_1}{\Lambda^2}{\bar L}{\tilde \phi}
\sigma^{\mu\nu}\nu_R B_{\mu\nu}\ , \hspace{1.5cm}
{\cal O}^{(6)}_W  =  \frac{g_2}{\Lambda^2} {\bar L}\tau^a {\tilde \phi} 
\sigma^{\mu\nu}\nu_R W_{\mu\nu}^a\ . 
\end{equation}
After spontaneous symmetry breaking, both ${\cal O}^{(6)}_B$ and
${\cal O}^{(6)}_W$ contribute to the magnetic moment.  Through
renormalization, these operators will also generate a contribution to
the 6D neutrino mass operator
\begin{equation}
{\cal O}^{(6)}_M =  \frac{1}{\Lambda^2}{\bar L}{\tilde \phi}\nu_R \left(\phi^\dag\phi\right) \ . 
\end{equation}
The three operators, $ \left\{ {\cal O}^{(6)}_B, {\cal O}^{(6)}_W,
{\cal O}^{(6)}_M \right\}$ constitute a closed set under
renormalization, so that our effective Lagrangian is given by
\begin{equation}
{\cal L}_{\rm eff}= C_B(\mu) {\cal O}^{(6)}_B + 
C_W(\mu) {\cal O}^{(6)}_W + C_M(\mu) {\cal O}^{(6)}_M\ ,
\end{equation}
where the operator coefficients, $C_i(\mu)$, depend upon the energy scale $\mu$.
The magnetic moment and mass are related to the operator coefficients as
\begin{eqnarray}
\frac{\mu_\nu}{\mu_B} &=&
-4\sqrt{2}\left( \frac{m_e v}{\Lambda^2}\right) \left[C_B(v)+C_W(v)\right]\ ,\\
\delta m_\nu &=& -C_M(v) \frac{v^3}{2\sqrt{2}\Lambda^2}\ .
\end{eqnarray}
To connect $\mu_\nu$ with $m_\nu$, we thus need to find the
relationship between the coefficients $C_i(\mu)$ at the weak scale,
$\mu=v$.  This requires that we solve the renormalization group
equations (RGE) which relate $C_i(\Lambda)$ to $C_j(v)$.

\begin{figure}
\label{oneloopa}
   \includegraphics[width=4cm]{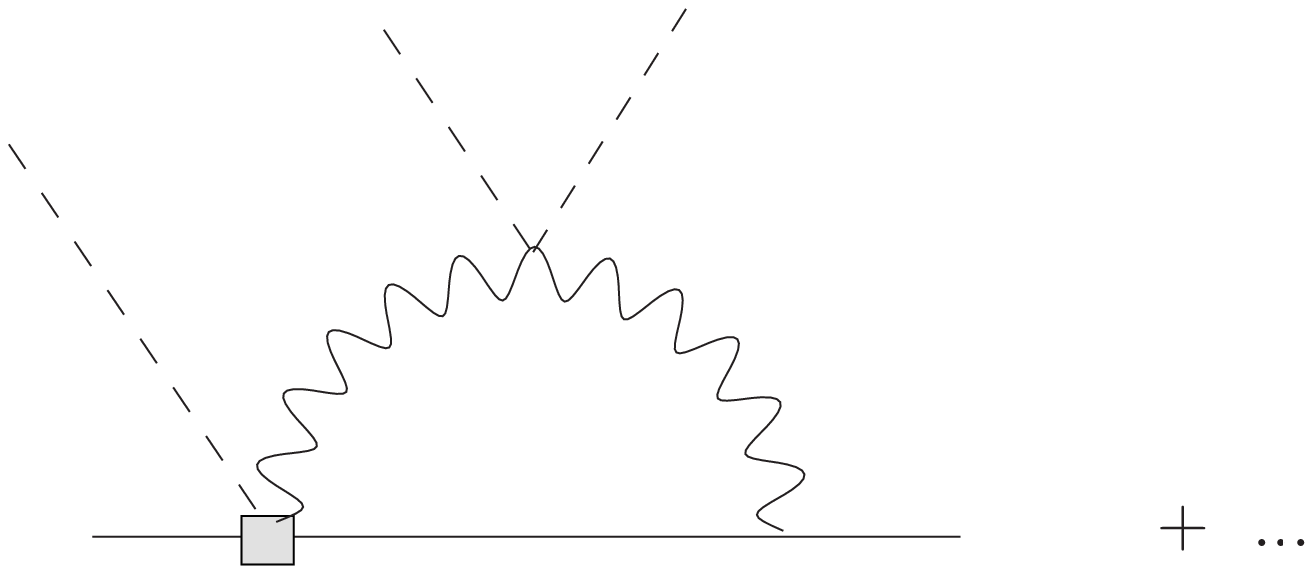} \hspace{4cm}
   \includegraphics[width=3.5cm]{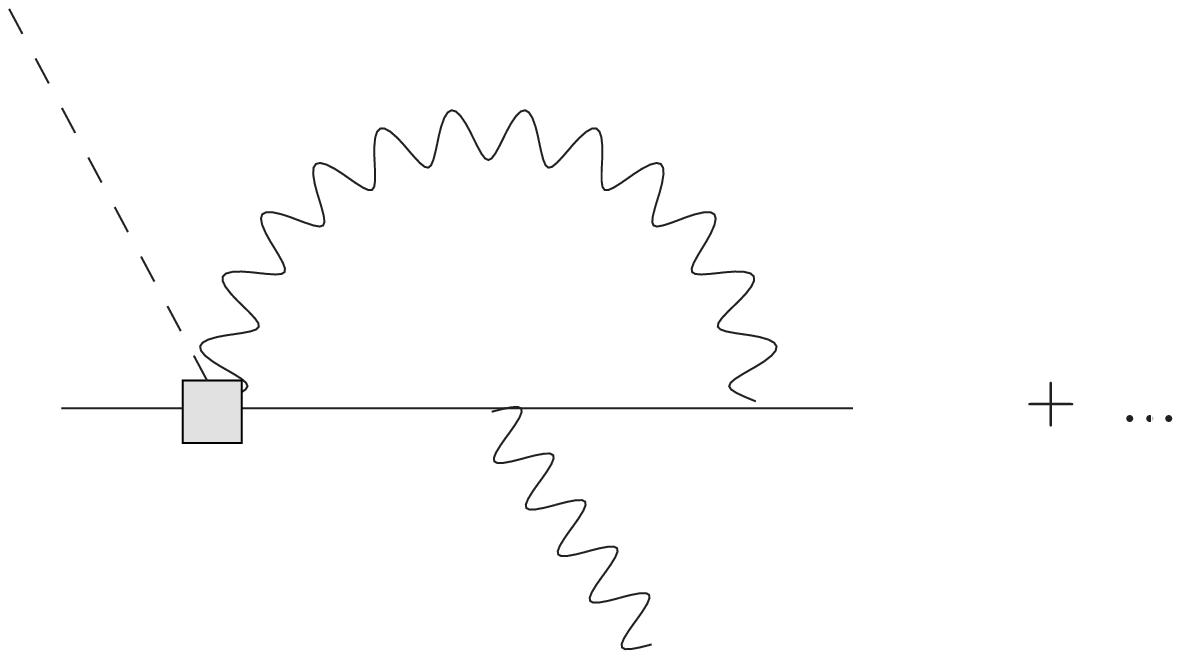} 
\caption{Renormalization of the mass operator, ${\cal O}^{(6)}_M$, due to insertions of ${\cal O}^{(6)}_{B,W}$ (left); self-renormalization of  ${\cal O}^{(6)}_{B,W}$ (right).} 
\end{figure}

\begin{figure}
\label{oneloopb}
   \includegraphics[width=7cm]{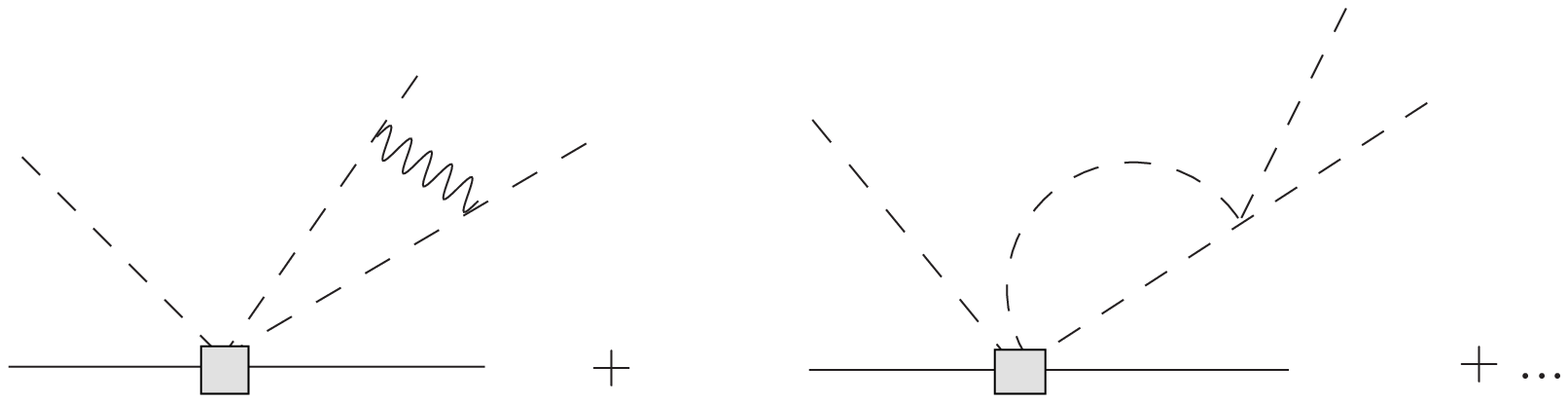} 
\caption{Self-renormalization of  ${\cal O}^{(6)}_{M}$.}
\end{figure}

Figures~(\ref{oneloopa},\ref{oneloopb}) display representative examples of
the one-loop diagrams which contribute to the renormalization of the
6D operators.  
(See Ref.~\cite{Bell} for further details.)
Solving the RGE, retaining only the leading
logarithms, we find that $\mu_\nu$ and $m_\nu$ are related as
\begin{equation}
\label{eq:massmurel}
\frac{|\mu_\nu|}{\mu_B} = 
\frac{G_F \, m_e }{\sqrt{2}\pi \alpha }\ 
\left[ \frac{\delta m_\nu}{  \alpha \, \ln ( \Lambda/v) } \right]  
\ \frac{32\pi\sin^4\theta_W}{9 \,  |f|} \ , 
\end{equation}
where $\theta_W$ is the weak mixing angle,
\begin{equation}
\label{f}
f=\left(1-r \right)-\frac{2}{3} r \tan^2\theta_W
-\frac{1}{3}\left(1+r\right)\tan^4\theta_W \ ,
\end{equation}
and $r$ is a ratio of operator coefficients at scale $\Lambda$, 
$r \equiv \left[ C_B(\Lambda)- C_W(\Lambda)\right]/
\left[ C_B(\Lambda)+ C_W(\Lambda)\right]$.

For $\Lambda \gsim 1~{\rm TeV}$, the bound becomes
\begin{equation}
 \label{eq:massbound}
 \frac{|\mu_\nu|}{\mu_B} \lsim 8\times 10^{-15}\times \left(\frac{\delta m_\nu}{1\ {\rm eV}}\right) \frac{1}{|f|} \ ,
 \end{equation}
and so for $f \simeq 1$ and $m_\nu \lsim 1$ eV, we find $|\mu_\nu| \lsim
10^{-14}\mu_B$.  In principle, larger values of $\mu_\nu$ could be
obtained, but only by fine-tuning the coefficients $C_i(\Lambda)$
to arrange cancellations in Eq.~\ref{f} such that $f\ll 1$.  
We therefore conclude that the natural size of $\mu_\nu$ for 
Dirac neutrinos is at least 
$10^2$ times
stronger than
astrophysical limits and $10^4$ times stronger than reactor
and solar neutrino bounds. 
The limits that can be placed on transition magnetic moments of
Majorana neutrinos are substantially weaker than the Dirac case, and
have recently been calculated in~\cite{Davidson}.



\vspace{-2mm}

\begin{theacknowledgments}
\vspace{-2mm}
This work was supported in part under U.S. DOE contracts
DE-FG02-05ER41361 and DE-FG03-92ER40701, and NSF grant PHY-0071856.

\end{theacknowledgments}

\end{document}